\documentclass[aps, prd,twocolumn, showpacs, showkeys,nofootinbib,floatfix]{revtex4-1}

\usepackage{amssymb}
\usepackage{amsmath}
\usepackage{graphicx}
\usepackage{hyperref}

\begin{document}

\title{Growth Index after the {\it Planck} Results}

\author{Lixin Xu}
\email{lxxu@dlut.edu.cn}

\affiliation{Institute of Theoretical Physics, School of Physics \&
Optoelectronic Technology, Dalian University of Technology, Dalian,
116024, P. R. China}

\affiliation{College of Advanced Science \& Technology, Dalian
University of Technology, Dalian, 116024, P. R. China}

\affiliation{Kavli Institute for Theoretical Physics China, CAS, Beijing, 100190, P. R. China}

\begin{abstract}
The growth index $\gamma_L$ was proposed to investigate the possible deviation from the standard $\Lambda$CDM model and Einstein's gravity theory in a dynamical perspective. Recently, thanks to the measurement of the cosmic growth rate via the redshift-space distortion, one can understand the evolution of density contrast through $f\sigma_8(z)$, where $f(z)=d\ln \delta/d \ln a$ is the growth rate of matter and $\sigma_8(z)$ is the rms amplitude of the density contrast $\delta$ at the comoving $8h^{-1}$ Mpc scale. In this paper, we use the redshift space distortion data points to study the growth index on the bases of Einstein's gravity theory and a modified gravity theory under the assumption of $f=\Omega_m(a)^{\gamma_L}$. The cosmic background evolution is fixed by the cosmic observations from the type Ia supernovae SNLS3, cosmic microwave background radiation data from {\it Planck} and baryon acoustic oscillations. Via the Markov Chain Monte Carlo method, we found the $\gamma_L$ values for Einstein's gravity with a cosmological constant, $w=constant$ dark energy and a modified gravity theory in the $1,2,3\sigma$ regions respectively: $0.675_{-0.0662-0.120-0.155}^{+0.0611+0.129+0.178}$, $0.745_{-0.0819-0.146-0.190}^{+0.0755+0.157+0.205}$ and $0.555_{-0.0167-0.0373-0.0516}^{+0.0193+0.0335+0.0436}$. In the Einstein's gravity theory, the values of growth index $\gamma_L$ show almost $2\sigma$ deviation from the theoretical prediction $6/11$ for the $\Lambda$CDM model. However in the modified gravity framework, a deviation from the Einstein's relativity is not detected in $1\sigma$ region. That implies that the currently available cosmic observations don't expect an alternative modified gravity theory beyond the $\Lambda$CDM model under Einstein's gravity, but that the simple assumption of $f=\Omega_m^{\gamma_L}$ should be improved.	 

\end{abstract}



\maketitle

\section{Introduction}

The data from the observations of type Ia supernovae (SNIa) implies that our Universe is undergoing an accelerated expansion during the late cosmic time \cite{ref:Riess98,ref:Perlmuter99}. To explain this accelerated expansion phase, dark energy (DE) in general relativity (GR) or a modified gravity (MG) theory at large scales was introduced, for the reviews about dark energy please see   
\cite{ref:DEReview1,ref:DEReview2,ref:DEReview3,ref:DEReview4,ref:DEReview5,ref:DEReview6,ref:DEReview7} and see \cite{ref:MG1,ref:MG2} for MG theory. Both of them can describe the same accelerated expansion phase, but they are different in nature. The former implies a new unknown energy component in our Universe, while the later means a new theory of gravity. Therefore it is important to discriminate one from the other. 

As is well known, for a given MG model an effective dark energy can be derived at the background level. It means that, for a given DE model in GR, one can find a corresponding MG model which can mimic the whole evolution history of DE model in GR at the background level. It is therefore impossible to discriminate MG theory from GR theory at the background level \cite{ref:background}, and as such the possibility using the geometric observations to discriminate the MG theory from GR is terminated.   

To discriminate MG theory from GR, large scale structure information should be employed. In recent decades the growth index is advocated to discriminate the dark energy models and modified gravity theory in the last decades \cite{Peebles,Fry,Lightman,wang,lue,Aquaviva,gong08b,polarski,linder,Koyama,Koivisto,Daniel,knox,ishak2006,laszlo,Zhang,Hu,ref:Ishak}. As a starting point, the matter perturbation in the subhorizon ($k\gg aH$) satisfies the following equation in the gauge-invariant comoving density contrast $\Delta \equiv \delta+3aHv/k$ \cite{ref:Pogosian2010}
\begin{equation}
\Delta^{''}+\left[2+\frac{H'}{H}\right]\Delta'-\frac{3}{2}\frac{E_m}{E}\mu\Delta=0, \label{eq:evolution}
\end{equation}
where $\delta=\delta\rho/\rho$ is the density contrast, $H=\dot{a}/a$ is the Hubble parameter, $E_m=\Omega_{m}a^{-3}$ and $E=H^2/H^2_0$. Here the  $\dot{}$ denotes a derivative with respect to the cosmic time $t$, and the $^{'}$ denotes a derivative with respect to $\ln a$. The function $\mu(k,a)$ comes from a parametrized modification to the Poisson equation \cite{ref:MGCAMB}
\begin{equation}
k^2\Psi=-4\pi G a^2\mu(a,k)\left[\rho\Delta+3(\rho+P)\sigma\right],
\end{equation}
due to a possible modification to the gravity theory, where $\sigma$ is the anisotropic stress. Here $\Psi$ is one of the scalar perturbations of the line element in the conformal Newtonian gauge
\begin{equation}
ds^2=-a^2(\tau)\left[(1+2\Psi)d\tau^2-(1-2\Phi)d\overrightarrow{x}^2\right],
\end{equation}
and the relation between $\Psi$ and $\Phi$ is parametrized as \cite{ref:MGCAMB}
\begin{equation}
k^2\left[\Phi-\gamma(k,a)\Psi\right]=\mu(k,a)12\pi G a^2(\rho+P)\sigma.
\end{equation}
Via the definition of the gauge-invariant
growth factor $f=d\ln\Delta/d\ln a$, which is reduced to $f=d\ln\delta/d\ln a$ in the matter comoving gauge, the perturbation evolution equation (\ref{eq:evolution}) can be rewritten as
\begin{equation}
f'+f^2+\left[2+\frac{H'}{H}\right]=\frac{3\mu}{2}\Omega_m(a)\label{eq:f}
\end{equation}
where $\Omega_{m}(a)=H^2_0\Omega_{m}a^{-3}/H^2$ is the dimensionless dark matter energy density. One can clearly see that the general function $\mu$ effectively corresponds to a modification to the Newtonian constant $G$ via the relation 
\begin{equation}
\mu=G_{eff}/G,
\end{equation}
where $G_{eff}$ is an effective Newtonian constant. In GR, $\mu\equiv 1$ is respected.

In general, analytical solutions to Eq. (\ref{eq:f}) are difficult
to obtain, and numerical methods are used. However, a good
approximation to the growth factor $f$ is given in the form of
\cite{ref:Pogosian2010,ref:growthindex}
\begin{equation}
f\equiv \frac{d\ln \Delta}{d\ln a}=
\Omega_{m}(a)^{\gamma_L}\label{eq:GI}
\end{equation}
where $\gamma_L$ is the growth index which is a constant in general, and the subscript $_L$ is added to distinguish it from the function $\gamma(k,a)$. Actually
the possible running of this index was also considered; for
the recent progress on a parametrized approach, please see
\cite{ref:recentgi}. The focus of this paper is to investigate the growth index in GR and MG theory. We will therefore take $\gamma_L$ as a constant. The running of the growth index will be reported in another work \cite{ref:Xugrowthruning}.   

In the literature, the growth rate of structure $f$  has been used to constrain dark energy models and to investigate the growth index including its running; see \cite{ref:weakness} for an example. However, the observed values of the growth rate $f_{obs}=\beta b$ are derived from the redshift space distortion parameter $\beta(z)$ and the linear bias $b(z)$, where a particular fiducial $\Lambda$CDM model is used. This means that the current $f_{obs}$ data can only be used to test the consistency of $\Lambda$CDM model. This is the weak point of using $f_{obs}$ data points. Moreover, the measurements of the linear growth rate are degenerate with the bias $b$ or clustering amplitude in the power spectra. To remove this weakness, Song and Percival proposed to use $f\sigma_8(z)$, which is almost model-independent and provides a good test for dark energy models even without the knowledge of the bias or $\sigma_8$ \cite{ref:Song}. Recently, the authors of Ref. \cite{ref:fsigma8total-Samushia2013} compiled the redshift space distortion (RSD) based $f\sigma_8$ measurement, which amounts to eight $f\sigma_8$ data points in the redshift range $z\in [0.17,0.78]$. Another data point at $z=0.80$ was also obtained in Ref. \cite{ref:fsigma86-Torre2013}. All of them allow us to {\it reconsider} the constraint on the growth index. For convenience, the data points are summarized in Table \ref{tab:fsigma8data}. Actually, the growth index was obtained for $\Lambda$CDM model in the constant case $\gamma_L=0.64\pm 0.05$ (assuming $f=\Omega_m(a)^{\gamma_{L}}$) in \cite{ref:fsigma8total-Samushia2013}. The $f\sigma_8(z)$ data points have been used to constrain dark energy model; see, for a example, \cite{ref:RSDXu} . 
\begin{center}
\begin{table}[tbh]
\begin{tabular}{ccl}
\hline\hline z & $f\sigma_8(z)$ & Survey and Refs \\ \hline
$0.067$ & $0.42\pm0.06$ & 6dFGRS~(2012) \cite{ref:fsigma85-Reid2012}\\
$0.17$ & $0.51\pm0.06$ & 2dFGRS~(2004) \cite{ref:fsigma81-Percival2004}\\
$0.22$ & $0.42\pm0.07$ & WiggleZ~(2011) \cite{ref:fsigma82-Blake2011}\\
$0.25$ & $0.39\pm0.05$ & SDSS~LRG~(2011) \cite{ref:fsigma83-Samushia2012}\\
$0.37$ & $0.43\pm0.04$ & SDSS~LRG~(2011) \cite{ref:fsigma83-Samushia2012}\\
$0.41$ & $0.45\pm0.04$ & WiggleZ~(2011) \cite{ref:fsigma82-Blake2011}\\
$0.57$ & $0.43\pm0.03$ & BOSS~CMASS~(2012) \cite{ref:fsigma84-Reid2012}\\
$0.60$ & $0.43\pm0.04$ & WiggleZ~(2011) \cite{ref:fsigma82-Blake2011}\\
$0.78$ & $0.38\pm0.04$ & WiggleZ~(2011) \cite{ref:fsigma82-Blake2011}\\
$0.80$ & $0.47\pm0.08$ & VIPERS~(2013) \cite{ref:fsigma86-Torre2013}\\
\hline\hline
\end{tabular}
\caption{The data points of $f\sigma_8(z)$ measured from RSD with the survey references.}
\label{tab:fsigma8data}
\end{table}
\end{center}

In March 2013, the European Space Agency (ESA) and the {\it Planck} Collaboration publicly released the first 15.5 months of cosmic microwave background (CMB) data which improves the constraint on the cosmological model \cite{ref:Planck}. So it is a good time to combine the RSD and {\it Planck} data points to investigate the growth index based on Einstein's general relativity and a modified gravity theory.

Firstly, we investigate the growth index in the framework of GR, where the $\Lambda$CDM model and $w=const$ dark energy model are considered. In the $\Lambda$CDM model, a deviation of the $\gamma_L$ value from the theoretical prediction up to the $2\sigma$ confidence level is found. This is consistent with the result obtained in Ref. \cite{ref:fsigma8total-Samushia2013} ($\gamma_L=0.64\pm 0.05$). It confirms the reliability of our code. One may think that GR is correct, but this discrepancy is due to an unsuitable assumption of a cosmological constant $\Lambda$. So one expects that the theoretical values are consistent to observed values for a dark energy model beyond the cosmological-constant one. As a simple extension to the $\Lambda$CDM model, the growth index for the $w$CDM model is investigated. We obtain the result $\gamma_L=0.745_{-0.0819-0.146-0.190}^{+0.0755+0.157+0.205}$ which deviates from the theoretical value $\gamma_L=3(w-1)/(6w-5)=0.543$ up to the $3\sigma$ confidence level. It implies that the dark energy model cannot alleviate the deviation. On the other hand, we consider the growth index in a modified gravity theory where an effective time-variable Newtonian constant is introduced. Under this assumption, no significant deviation from GR was detected via the currently available cosmic observations. Based on these findings, one could conclude that $\Lambda$CDM in GR is still the simplest model that is consistent to the currently available cosmic observations, and that the discrepancy between the observed and theoretical predictions comes from an unsuitable assumption about the form of the growth function $f=\Omega_m^{\gamma_L}$ or a nonconstant growth index $\gamma_L$.

This paper is structured as follows. In Sec. \ref{sec:BE}, we present the basic background evolution equations and the relation between $\gamma_L$ and MG theory. The data points and constrained results are summarized in Sec. \ref{sec:datasets}. We give the conclusion in Sec. \ref{sec:con}.

\section{Basic Equations} \label{sec:BE}

For a spatially flat Friedmann-Robertson-Walker universe filled with matter, a cosmological constant and radiation, the
background evolution equation is written as
\begin{equation}
H^2=\frac{8\pi
G_{eff}}{3}\left(\rho_{r}+\rho_{m}+\rho_{\Lambda}\right)\label{eq:FE}
\end{equation}
where $\rho_{r}$, $\rho_{m}$ and $\rho_{\Lambda}$ are the energy density for the radiation, the matter
and the cosmological constant respectively. 

Once a form for the growth index $\gamma_{L}$ is given, from Eq. (\ref{eq:f}) one has \cite{ref:Pogosian2010}
\begin{equation}
\mu=\frac{2}{3}\Omega^{\gamma_L-1}_{m}\left[\Omega^{\gamma_L}_{m}+2+\frac{H'}{H}+\gamma_L\frac{\Omega^{'}_{m}}{\Omega_m}+\gamma^{'}_{L}\ln(\Omega_m)\right].
\end{equation}
The other parametrized modification to GR is set to $\gamma(k,a)=1$ in this case \cite{ref:Pogosian2010}. As was said in the Introduction, an MG theory can mimic the $\Lambda$CDM model in GR at the background level. So in the MG case, we still assume the same Hubble expansion rate as in Eq. (\ref{eq:FE}). 

\section{Data Sets and Results} \label{sec:datasets}

To study the evolution of the perturbation, we should fix the background evolution. To realize this, we use the cosmic observations from SNIa SNLS3 data, baryon acoustic oscillations (BAO), and the recently released {\it Planck} CMB data points. 

To use the supernovae as 'standard candles', the luminosity distances will be employed. We use the SNLS3 which consists of $472$ SN calibrated by SiFTO and SALT2; for the details please see \cite{ref:SNLS3}.

To use BAO as a 'standard ruler', we use the measured ratio of $D_V/r_s$, where $r_s$ is the comoving sound-horizon scale at the recombination epoch, and $D_V$ is the 'volume distance'
\begin{equation}
D_V(z)=[(1+z)^2D^2_A(z)cz/H(z)]^{1/3},
\end{equation}
where $D_A$ is the angular diameter distance. We use the BAO data points in the someway as in the {\it Planck} paper; for details please see Sec. 5.2 of Ref. \cite{ref:Planck}.  

To fix the other relevant cosmological model parameters, we use the recently released {\it Planck} data sets which include the high-l TT likelihood ({\it CAMSpec}) up to a maximum multipole number of $l_{max}=2500$ from $l=50$, the low-l TT likelihood ({\it lowl}) up to $l=49$ and the low-l TE, EE, BB likelihood up to $l=32$ from WMAP9, the data sets are available on line \cite{ref:Planckdata}. Then the amplitude of perturbations can be determined by the CMB. We will use  {\bf MGCAMB} \cite{ref:MGCAMB} which is a modified version based on the publicly available package {\bf CAMB} \cite{ref:CAMB} to calculate the evolutions of the perturbations and the CMB and matter power spectra. We first have modified the code to obtain the values of $\sigma_8$ at any redshift $z$. A new module was added to calculate the growth rate $f\sigma_8(z)$. Here $f(z)$ is given by Eq. (\ref{eq:GI}). Therefore, when the code is running, one can obtain the values of $f\sigma_8(z)$ at any redshift $z$.

The corresponding $\chi^2_{RSD}$ is given as
\begin{equation}
\chi^{2}_{RSD}=\sum^{N_{RSD}}_{i=1}\frac{(f\sigma_8(z_i)-f\sigma^{obs}_8(z_i))^2}{\sigma^2_{f\sigma_8(z_i)}}.
\end{equation}
In this paper, we will use the $N_{RSD}=10$ data points summarized in Table \ref{tab:fsigma8data} to constrain the model parameter space. 

To obtain the distribution of the model parameter space, we calculate the total likelihood $\mathcal{L} \propto e^{-\chi^{2}/2}$, where the $\chi^{2}$ is given as
\begin{equation}
\chi^{2}=\chi^{2}_{CMB}+\chi^{2}_{SN}+\chi^{2}_{BAO}+\chi^{2}_{RSD}.
\end{equation}
We perform a global fitting to the model parameter space by using the Markov Chain Monte Carlo (MCMC) method. We modified the publicly available {\bf cosmoMC} \cite{ref:MCMC} package to calculate $\chi^{2}_{RSD}$. 

The following seven- or eight-dimensional parameter space for the $\Lambda$CDM or $w$CDM model is adopted
\begin{equation}
P\equiv\{\omega_{b},\omega_c, 100\Theta_{MC},(\text{or~} w),\gamma_L,\tau,n_{s},\ln[10^{10}A_{s}]\}
\end{equation}
where $\omega_{b}=\Omega_{b}h^{2}$ and $\omega_{c}=\Omega_{c}h^{2}$ are the physical density of  the baryon and the cold dark matter respectively, $\Theta_{MC}$ (multiplied by $100$) is the ratio of the sound horizon and the angular diameter distance, $\tau$ is the optical depth, $\gamma_L$ is the newly added model parameter related to the growth index, $n_{s}$ is the scalar spectral index, and $A_{s}$ is the amplitude of the initial power spectrum. The pivot scale of the initial scalar power spectrum $k_{s0}=0.05\text{Mpc}^{-1}$ is used in this paper. The following priors to the model parameters are adopted: $\omega_{b}\in[0.005,0.1]$, $\omega_{c}\in[0.01,0.99]$, $\Theta_{MC}\in[0.5,10]$, $\tau\in[0.01,0.8]$, $w\in[-2,0]$, $\gamma_L\in[0,1]$,$n_{s}\in[0.5,1.5]$, $\log[10^{10}A_{s}]\in[2.7, 4]$. Furthermore, the hard coded prior on the comic age $10\text{Gyr}<t_{0}<\text{20Gyr}$ is also imposed. The new Hubble constant $H_{0}=73.8\pm2.4\text{kms}^{-1}\text{Mpc}^{-1}$ \cite{ref:hubble} is adopted.

\begin{center}
\begin{figure}[tbh]
\includegraphics[width=8cm]{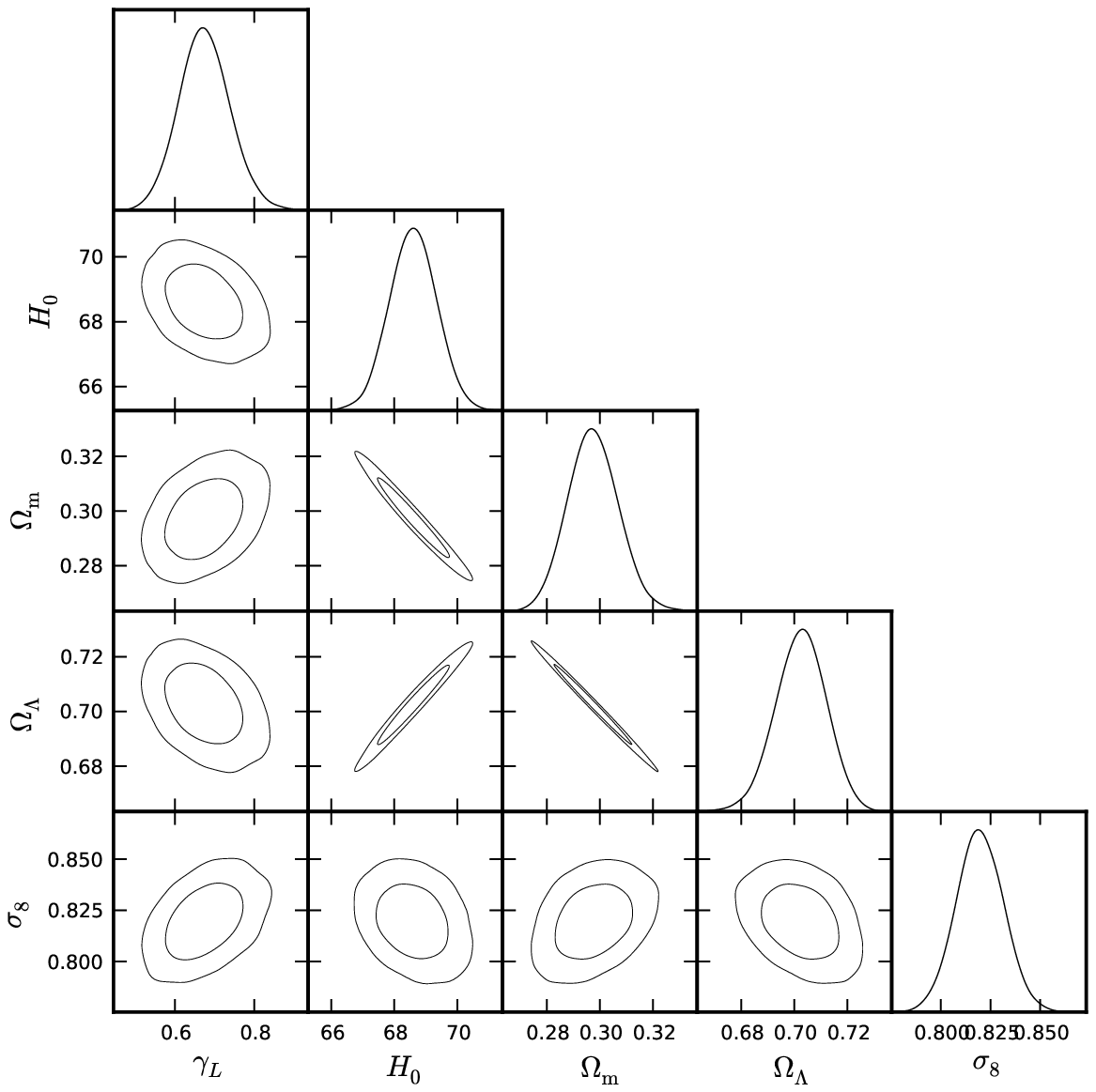}
\includegraphics[width=8cm]{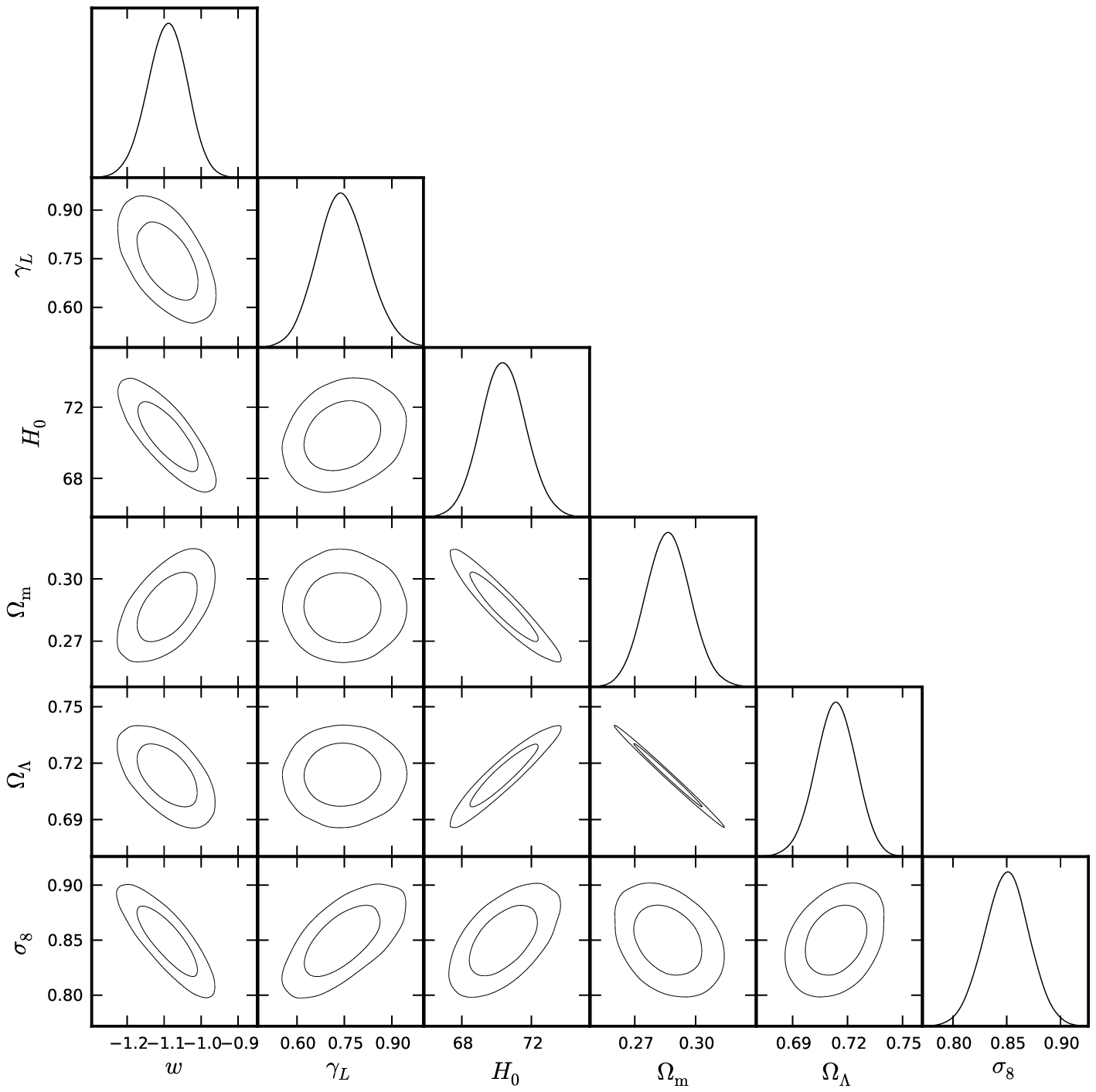}
\includegraphics[width=8cm]{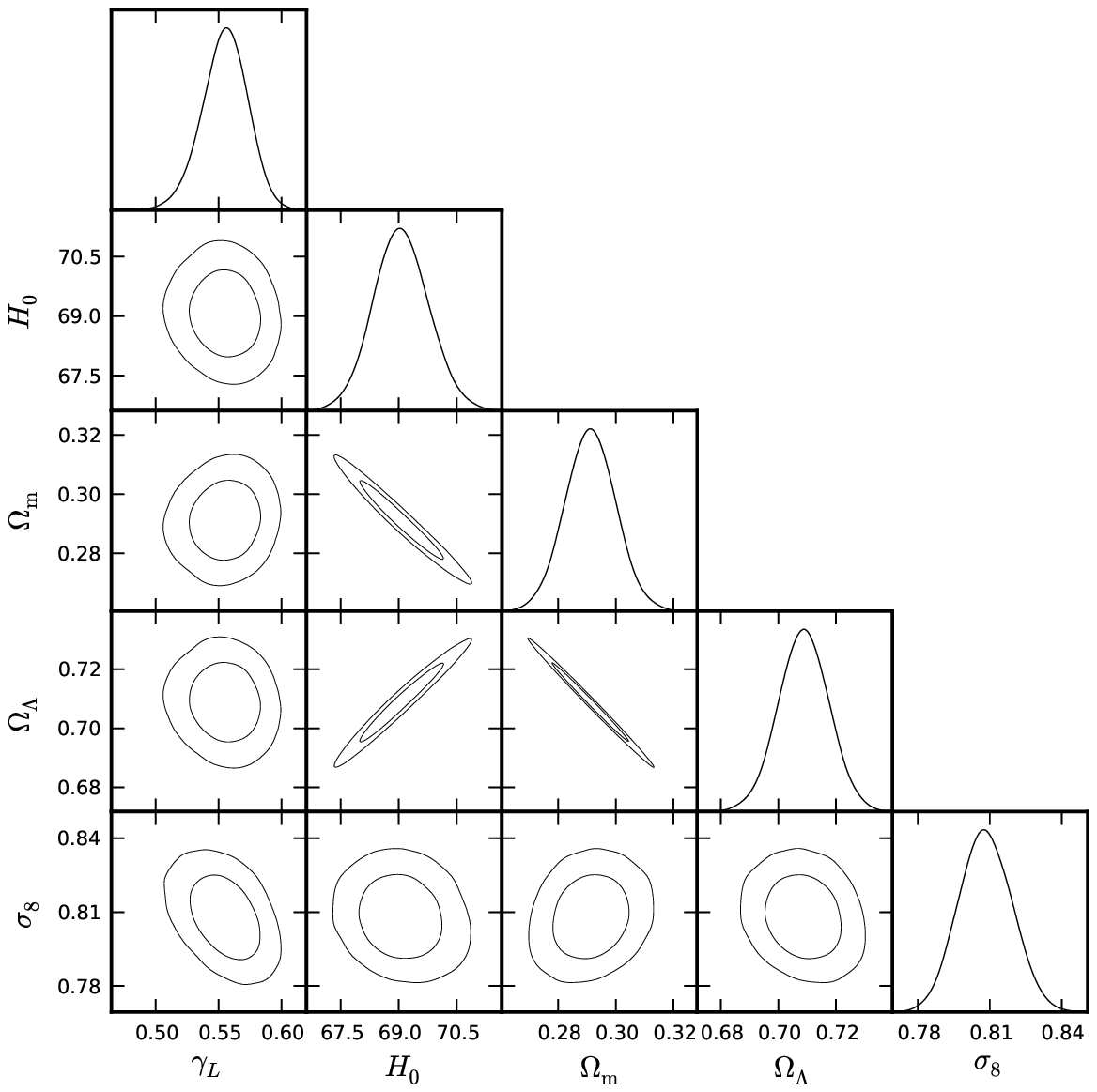}
\caption{The contour plots of interesting model parameters for GR (upper panels) and MG (bottom panel).}\label{fig:contour}
\end{figure}
\end{center}

\begin{center}
\begin{figure}[tbh]
\includegraphics[width=9cm]{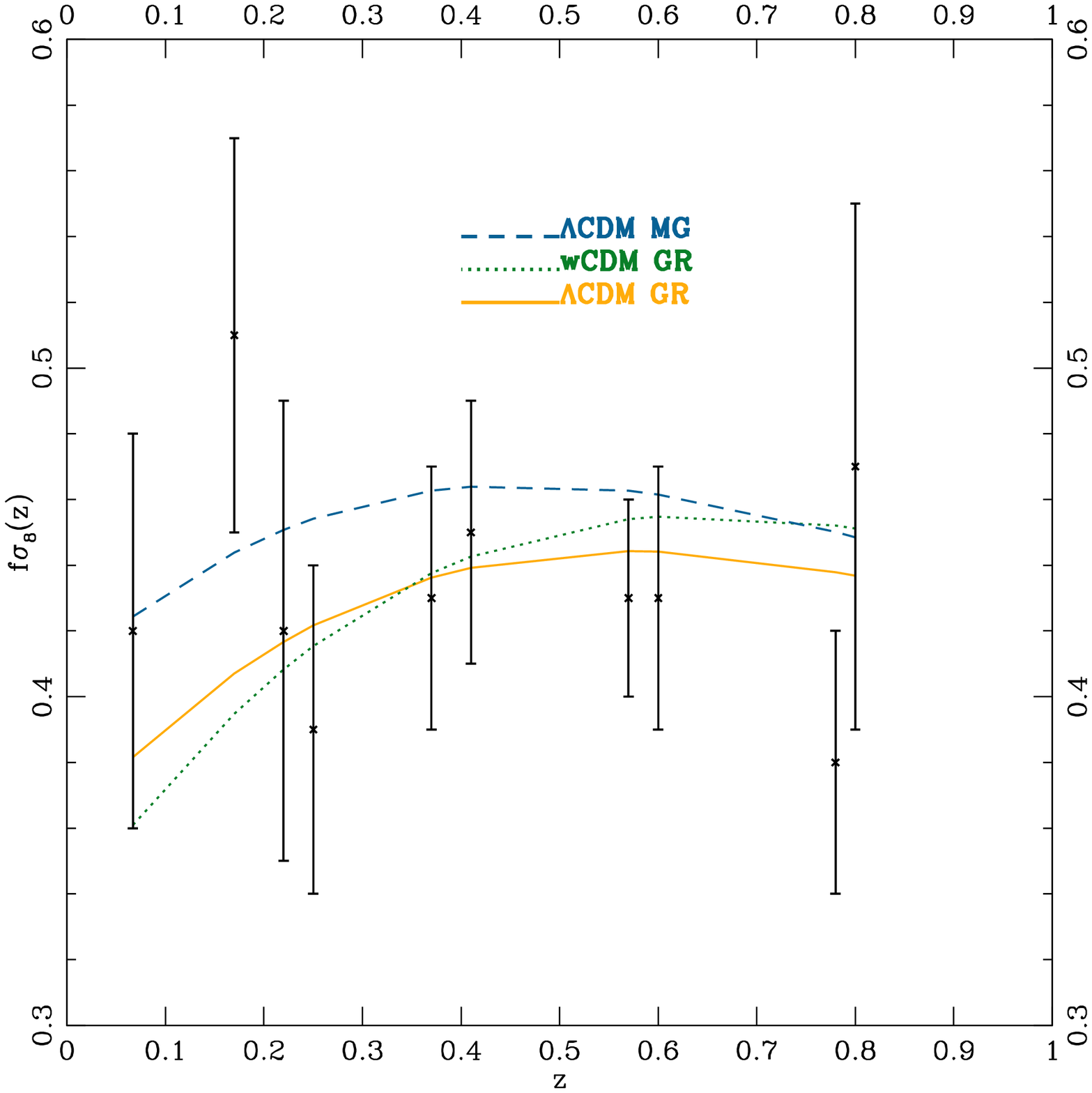}
\caption{$f\sigma_8(z)$ redshift $z$ for GR (the yellow solid curve) and MG (the dotted green curve) in the $\Lambda$CDM model, where the relevant cosmological parameters are fixed to their mean values as listed in Table \ref{tab:results}. The black lines with error bars are the RSD data as listed in Table \ref{tab:fsigma8data}.}\label{fig:fsigma8}
\end{figure}
\end{center}

We ran eight chains for every cosmological model with the growth index for GR and MG theory on the {\it Computational Cluster for Cosmos} and stopped sampling when the worst e-values [the variance(mean)/mean(variance) of 1/2 chains] $R-1$ were of the order $0.001$ which confirmes the convergence of the chains. The obtained results are shown in Table \ref{tab:results} for GR and MG theory. Also, the one-dimensional probability distribution and the two-dimensional contours for the cosmological interesting model parameters are plotted in Fig. \ref{fig:contour}. The evolutions of $f\sigma_8(z)$ with respect to the redshift $z$ is shown in Fig. \ref{fig:fsigma8}, where the mean values are adopted for GR and MG theory. Higher values of $f\sigma_8(z)$ are favored in MG theory. 

\begin{widetext}
\begingroup
\squeezetable
\begin{center}
\begin{table}[tbh]
\begin{tabular}{lllllll}
\hline\hline 
MP &Mean \& errors (GR) & BF (GR) & Mean \& errors (wCDM) & BF (wCDM) & Mean \& errors (MG) & BF (MG)\\ \hline
$\Omega_b h^2$ & $0.0223_{-0.000246-0.000481-0.000631}^{+0.000244+0.000496+0.000654}$ & $0.0222$ & $0.0221_{-0.000253-0.000488-0.000631}^{+0.000249+0.000507+0.000674}$ & $0.0220$ & $0.0223_{-0.000247-0.000475-0.000611}^{+0.000246+0.000494+0.000649}$ & $0.0223$\\ 
$\Omega_c h^2$ & $0.117_{-0.00161-0.00312-0.00395}^{+0.00159+0.00325+0.00445}$ & $0.118$ & $0.119_{-0.00194-0.00386-0.00507}^{+0.00194+0.00377+0.00498}$ & $0.118$ & $0.116_{-0.00148-0.00296-0.00385}^{+0.00148+0.00289+0.00387}$ & $0.116$\\
$100\theta_{MC}$ & $1.0417_{-0.000556-0.00111-0.00144}^{+0.000555+0.00107+0.00143}$ & $1.0414$ & $1.0414_{-0.000572-0.00113-0.00148}^{+0.000575+0.00113+0.00149}$ & $1.0413$ & $1.0418_{-0.000549-0.00111-0.00147}^{+0.000553+0.00110+0.00149}$ & $1.0418$\\ 
$\tau$ & $0.0921_{-0.0143-0.0249-0.0329}^{+0.0125+0.0268+0.0357}$ & $0.0940$ & $0.0870_{-0.0134-0.0238-0.0304}^{+0.0123+0.0257+0.0353}$ & $0.0836$ & $0.0857_{-0.0129-0.0224-0.0296}^{+0.0114+0.0248+0.0334}$ & $0.0868$\\
$w$ & $-$ & $-$ & $-1.0916_{-0.0518-0.105-0.139}^{+0.0521+0.100+0.132}$ & $-1.122$ & $-$ & $-$ \\
$\gamma_L$ & $0.675_{-0.0662-0.120-0.155}^{+0.0611+0.129+0.178}$ & $0.685$ & $0.745_{-0.0819-0.146-0.190}^{+0.0755+0.157+0.205}$ & $0.714$ & $0.555_{-0.0167-0.0373-0.0516}^{+0.0193+0.0335+0.0436}$ & $0.558$\\
$n_s$ & $0.967_{-0.00556-0.0112-0.0145}^{+0.00558+0.0110+0.0146}$ & $0.965$ & $0.962_{-0.00603-0.0120-0.0158}^{+0.00609+0.0121+0.0159}$ & $0.964$ & $0.969_{-0.00550-0.0110-0.0143}^{+0.00549+0.0108+0.0141}$ & $0.970$\\
${\rm{ln}}(10^{10} A_s)$ & $3.0876_{-0.0277-0.0492-0.0659}^{+0.0251+0.0521+0.0691}$ & $3.0926$ & $3.0823_{-0.0243-0.0471-0.0600}^{+0.0243+0.0496+0.0676}$ & $3.0766$ & $3.0719_{-0.0251-0.0437-0.0573}^{+0.0222+0.0479+0.0642}$ & $3.0772$\\
$\Omega_\Lambda$ & $0.702_{-0.00925-0.0190-0.0265}^{+0.00934+0.0179+0.0229}$ & $0.697$ & $0.713_{-0.0106-0.0216-0.0288}^{+0.0107+0.0205+0.0263}$ & $0.724$ & $0.709_{-0.00848-0.0171-0.0226}^{+0.00853+0.0169+0.0217}$ & $0.708$\\
$\Omega_m$ & $0.298_{-0.00934-0.0179-0.0229}^{+0.00925+0.0190+0.0265}$ & $0.303$ & $0.287_{-0.0107-0.0205-0.0263}^{+0.0106+0.0216+0.0288}$ & $0.276$ & $0.291_{-0.00853-0.0169-0.0217}^{+0.00848+0.0171+0.0227}$ & $0.292$\\
$\sigma_8$ & $0.820_{-0.0116-0.0232-0.0298}^{+0.0116+0.0233+0.0319}$ & $0.824$ & $0.850_{-0.0203-0.0403-0.0535}^{+0.0201+0.0398+0.0516}$ & $0.855$ & $0.808_{-0.0109-0.0208-0.0270}^{+0.0109+0.0214+0.0284}$ & $0.811$\\
$z_{re}$ & $11.155_{-1.0976-2.187-2.990}^{+1.109+2.192+2.840}$ & $11.385$ & $10.797_{-1.0777-2.145-2.798}^{+1.0822+2.135+2.849}$ & $10.518$ & $10.553_{-1.0105-1.993-2.699}^{+1.0181+2.0442+2.704}$ & $10.680$\\ 
$H_0$ & $68.598_{-0.722-1.437-1.978}^{+0.726+1.439+1.924}$ & $68.127$ & $70.395_{-1.241-2.422-3.190}^{+1.244+2.518+3.295}$ & $71.535$ & $69.0660_{-0.739-1.358-1.771}^{+0.685+1.412+1.861}$ & $69.00146$\\
${\rm{Age}}/{\rm{Gyr}}$ & $13.771_{-0.0355-0.0727-0.0956}^{+0.0360+0.0715+0.0952}$ & $13.793$& $13.752_{-0.0366-0.0731-0.0956}^{+0.0367+0.0718+0.0922}$ & $13.739$ & $13.760_{-0.0361-0.0725-0.0955}^{+0.0361+0.0716+0.0922}$ & $13.757$\\
\hline\hline
\end{tabular}
\caption{The mean values with $1,2,3\sigma$ errors and the best-fit values of model parameters for general relativity and modified gravity theory, where SNLS3, BAO, {\it Planck}+WMAP9 and RSD data sets are used.}\label{tab:results}
\end{table}
\end{center}
\endgroup
\end{widetext}

As shown in the second column of Table \ref{tab:results}, the resultant $\gamma_L=0.675_{-0.0662-0.120-0.155}^{+0.0611+0.129+0.178}$ for the $\Lambda$CDM model in GR is consistent with that obtained in Ref. \cite{ref:fsigma8total-Samushia2013} and Ref. \cite{ref:planckgrowth} where $\gamma_L$ is $0.64\pm 0.05$ with WMAP7 and $0.740\pm 0.14$ with {\it Planck} at the $1\sigma$ confidence level respectively. However, our result is somewhat greater and is almost $2\sigma$ away from the theoretical prediction $\gamma_\Lambda \approx 6/11$ (the theoretical prediction value of the growth index for the $\Lambda$CDM model in GR). But we should note that our result is consistent with other fitted values of $\gamma_L$ reported by different authors, such as, for example, Refs. \cite{gong08b,ref:gamma,ref:fsigma8total-Samushia2013,ref:weakness,ref:planckgrowth}. This consistence shows the reliability of our code and implies a possible deviation from the $\Lambda$CDM model at the $2\sigma$ level. 

One may consider that the possible discrepancy is due to the cosmological constant, i.e. a nonproper dark energy model, and that another dark energy model beyond a cosmological constant can alleviate this difficulty. However as is shown in the fourth column of Table \ref{tab:results}, the introduction of a $w=const$ dark energy model cannot give a consistent theoretical and cosmic observational values.  

Based on the above observations, one could conclude that GR+dark energy would not be a correct theory if one insists that the assumption $f=\Omega_m^{\gamma_L}$ is correct. Of course, this discrepancy may be due to the incorrect assumption of $f=\Omega_m^{\gamma_L}$, so one needs another test that is a possible modification of GR.  

\begin{center}
\begin{figure}[tbh]
\includegraphics[width=9cm]{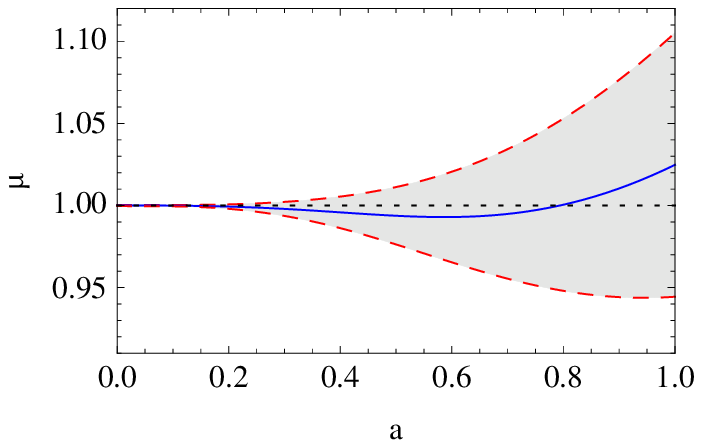}
\caption{The evolutions of $\mu$ with respect to the scale factor $a$ for MG (the blue curved line) in $1\sigma$ region bounded by the red dashed curves, where the black dotted straight line is for GR.}\label{fig:mu}
\end{figure}
\end{center}

As described in the Introduction, this parametrized $f=\Omega(a)^{\gamma_L}$ gives an effective modification to the gravity theory effectively, i.e., an effective Newtonian constant scaling to $G$. As a result, the Poisson equation will be modified. Then one can understand the effect of $\gamma_L$ to the CMB power spectrum on large scales. On the contrary, a tighter constraint to the model parameter $\gamma_L$ can be expected. When we consider the possible modification to GR due to the introduction of an effective time-variable Newtonian constant, the value of $\gamma_L$ is decreased to $0.555_{-0.0167-0.0373-0.0516}^{+0.0193+0.0335+0.0436}$ as shown in the sixth column in Table \ref{tab:results}. In the MG case, as expected a tighter constrain on the parameter $\gamma_L$ due to the possible modification to the Poisson equation is confirmed. From the contour plots of $\sigma_8$-$\gamma_L$, one can see that they are correlated in GR and anti-correlated in MG. It means that CMB can give a stronger constraint to $\gamma_L$ than that of RSD when the gravity deviates from GR in the case of an effective Newtonian constant. To see the deviation from GR due to the introduction of an effective time-variable Newtonian constant, we show the evolution of $\mu$ with respect to the scale factor $a$ in Fig. \ref{fig:mu}. To calculate the $1\sigma$ region, we consider the propagation of the errors for $\mu(a)$ and marginalize the other irrelevant model parameters by the Fisher matrix analysis \cite{ref:NRP,ref:Alam}. If the other 
irrelevant model parameters are not marginalized, the error bars will be underestimated. The errors are calculated by using the covariance matrix $C_{ij}$ of the fitting model parameters which is an output of {\bf cosmoMC}. The errors for a function $f=f(\theta)$ in terms of the variables $\theta$ are given via the formula \cite{ref:Alam,ref:Nesseris2005,ref:Wang2010}
\begin{equation}
\sigma^2_f=\sum^n_i\left(\frac{\partial f}{\partial \theta_i}\right)^2C_{ii}+2\sum^n_i\sum^n_{j=i+1}\left(\frac{\partial f}{\partial \theta_i}\right)\left(\frac{\partial f}{\partial \theta_j}\right)C_{ij}
\end{equation}
 where $n$ is the number of variables. For our case, $f$ is $\mu$ and $\theta=\{\Omega_b h^2,\Omega_c h^2,\gamma_L\}$. Then the $1\sigma$ errors for $\mu(a)$ are given by
 \begin{equation}
 \mu_{1\sigma}(z)=\mu(a)|_{\theta=\bar{\theta}}\pm\sigma_\mu,
 \end{equation}
where $\bar{\theta}$ are the mean values of the relevant model parameters as shown in Table \ref{tab:results}. In Fig. \ref{fig:mu} the shadows bounded by the colored curves denote the $1\sigma$ region of $\mu$. It is clear that one cannot detect the modification due to an effective time-variable Newtonian constant to GR in the $1\sigma$ region.    

\section{Conclusion} \label{sec:con}

In this paper, we studied the growth index in the framework of Einstein's general relativity and a modified gravity theory via the currently available cosmic observations which include SN, BAO, CMB, and RSDdata. The background evolution of the Universe is fixed by the standard candles, standard rulers, and the CMB. The overdensity evolution or the growth index is constrained by the RSD. In the GR cases for the $\Lambda$CDM and $w$CDM models, we found significant discrepancies (up to $2\sigma$) between the theoretical predictions and observed values of the growth index $\gamma_L$. One could then conclude that GR+dark energy would not be a correct theory or the assumption $f=\Omega_{m}^{\gamma_L}$ is incorrect. To confirm this we consider a possible modification to GR in terms of an effective time variable Newton constant. In the MG case, we did not find any deviation from GR in the $1\sigma$ region. As a final step, we report the nonproper assumption of $f=\Omega_{m}^{\gamma_L}$ with a constant $\gamma_L$ in GR for the $\Lambda$CDM and $w$CDM models. Maybe the running of the growth index or the mismatch between the theoretical and observed values of $\sigma_8$ can give a reasonable explanation for this discrepancy. This still deserves research in the future.   

\acknowledgements{The author thanks an anonymous referee for helpful improvement of this paper. L. Xu's work is supported in part by NSFC under the Grants No. 11275035 and "the Fundamental Research Funds for the Central Universities" under the Grants No. DUT13LK01.}

\end{document}